\documentclass[intlimits,twoside,a4paper]{article}

\usepackage{amsmath,amssymb}
\usepackage{graphicx}
\usepackage{xcolor}
\usepackage[cp1251]{inputenc}

\usepackage{cmpj3}

\issue{2017}{20}{3}{33801}
\doinumber{10.5488/CMP.20.33801}

\title[Charge reversal and surface charge amplification using the modified Poisson-Boltzmann theory]
{Charge reversal and surface charge amplification in asymmetric valence restricted primitive
model planar electric double layers in the modified Poisson-Boltzmann theory\thanks{It is a pleasure to dedicate
this paper to the memory of Dr. J.P. Badiali.}}

\author[L.B. Bhuiyan, C.W. Outhwaite]{L.B. Bhuiyan\refaddr{label1}, C.W. Outhwaite\refaddr{label2}}

\addresses{
\addr{label1} Laboratory of Theoretical Physics, Department of
Physics, Box 70377, University of Puerto Rico,\\ San Juan, Puerto
Rico 00936-8377, USA
\addr{label2} Department of Applied Mathematics, University of Sheffield,
Sheffield S3 7RH, UK
}

\date{Received April 27, 2017, in final form May 31, 2017}

\begin{document}

\maketitle

\begin{abstract}

The modified Poisson-Boltzmann theory of the restricted primitive model double layer
is revisited and recast in a fresh, slightly broader perspective. Derivation of relevant
equations follow the techniques utilized in the earlier MPB4 and MPB5 formulations and
clarifies the relationship between these. The MPB4, MPB5, and a new formulation of the theory
are employed in an analysis of the structure and charge reversal phenomenon in asymmetric
2:1/1:2 valence electrolytes. Furthermore, polarization induced surface charge amplification is
studied in 3:1/1:3 systems. The results are compared to the corresponding Monte Carlo simulations.
The theories are seen to predict the ``exact'' simulation data to varying degrees of accuracy ranging
from qualitative to almost quantitative. The results from a new version of the theory are found to be
of comparable accuracy as the MPB5 results in many situations. However, in some cases involving low
electrolyte concentrations, theoretical artifacts in the form of un-physical ``shoulders'' in the
singlet ionic distribution functions are observed.

\pacs 82.45.Fk, 61.20.Qg, 82.45.Gj
\end{abstract}

\section{Introduction}
Electric double layers play a key role in a wide variety of phenomena, in the fields diverse as
colloids, physical and biological systems, and electrochemistry \cite{belloni,levin,grosberg,chertsvy,bazant,pilon,
fedorov,henderson}. Recent years have seen an increasing relevance of double layers in DNA
sequencing \cite{sotres,maekawa}, drug delivery \cite{honary1,honary2}, and \emph{green}
technology, viz., designing novel energy storage devices as in super-capacitors
(see for example, reference \cite{galinski}).
An electric double layer appears when a system of charged particles in a fluid
is in the neighborhood of a charged surface. Various models and theories have been
developed depending upon the physical system being treated. A simple model of an
electrolyte next to a plane charged surface is a basic model in many areas.
Gouy and Chapman \cite{gouy1,chapman} analysed
this model, based on the Poisson-Boltzmann (PB) equation, where the electrolyte was treated
as a system of point ions moving in a medium of constant permittivity.
Later on,  Stern \cite{stern} proposed to lower the permittivity in the region immediately
next to the planar electrode (Stern layer) in order to bring the computed double layer capacitance in line
with the experiments. At not too high surface charge and electrolyte concentration for univalent ions,
the Gouy-Chapman-Stern (GCS) theory describes reasonably well the
diffuse part of the electric double layer. However, double layer research
over the last four decades involving formal statistical mechanical theories, simulations,
and experiments has shown the shortcomings of the theory in
capturing many physical phenomena (see for example, the review \cite{henderson}).
As a case in point, the GCS theory cannot explain the two phenomena that are the
focus of this work, namely, charge reversal (CR) (or charge inversion or overcharging)
and surface charge amplification (SCA). In CR, the local density of the counterions
overcompensate the surface charge, while in SCA, in the absence of specific ion adsorption,
there is an apparent increase in the surface charge due to an excess of coions.

CR was indicated in the simulation work of Torrie and Valleau \cite{torrie1,torrie2},
Snook and van Megen \cite{snook} and in the modified Poisson-Boltzmann (MPB) and
hypernetted-chain/mean spherical approximation (HNC/MSA) theories \cite{carnie}. The important
feature of these works was allowing the ions to have size and accounting for the
approximations in the PB equation. There is now a large body of theoretical and
experimental work on charge reversal in colloidal systems due to its importance in
both equilibrium and non-equilibrium situations \cite{belloni,levin,greberg,quesada-perez1,
lyklema,lozada-cassou,martin-molina1,semenov,martin-molina2,barrios,gonzalez-tovar}.
Surface charge amplification is more counterintuitive than CR. It can occur when
the ions have asymmetry in size and charge \cite{jimenez-angeles,guerrero-garcia1,quesada-perez2,patra,wang1}.
Imaging can also promote SCA, even when the ions are of equal size \cite{guerrero-garcia2,wang2,wang3}.
Here, we discuss the MPB theory with imaging for equisized ions with 1:2/2:1, 1:3/3:1 electrolytes.

\section{Model and theory} \label{Model}

The electric double layer is modelled by a restricted primitive model electrolyte in the neighborhood of a uniformly charged plane electrode. The solute molecules are hard spheres of diameter $a$ with point charges at their center moving in a medium of constant relative  permittivity $\epsilon _{r} $, while the electrode has a constant relative  permittivity of $\epsilon _{\text w}$. The analysis presented here builds upon the earlier formulations of MPB4 \cite{outh1,outh2} and MPB5 \cite{outh3} versions of the theory
and clarifies the relation between them. The present approach is a bit more general since the MPB5
situation will be seen to follow as a special limiting case. The connections to the MPB4 are also explored.

At a perpendicular distance $x$ from the electrode into the solution, the mean electrostatic potential $\psi(x)$ satisfies Poisson's equation
\begin{equation}
\frac{\rd^{2}\psi}{\rd x^{2}}=-\frac{1}{\epsilon_{0}\epsilon_{r}}\sum_{s}e_{s}n_{s}g_{s}(x), \label{eq1}
\end{equation}
where the sum is over the ion species, $\epsilon _{0}$ is the vacuum permittivity, $e_{s}$ is the charge of an ion of species~$s$, $n_{s}$ and $g_{s}(x)$ are the mean number density and singlet distribution function for an ion of species $s$, respectively. Kirkwood's charging process \cite{kirkwood} for an ion $i$ at $x_{1}$ gives
\begin{equation}
g_i (x_1 )=\zeta_{i}(x_1)\exp\Bigg[-\beta e_i \psi(x_1)-\beta \int_{0}^{e_i}\lim_{r\rightarrow 0}(\phi_{i}^{*}-\phi_{i\text{b}}^{*})\rd e_i\Bigg], \label{eq2}
\end{equation}
where
\begin{equation}
\phi_{i}^{*}=\phi_{i}(1;2)-\frac{e_i}{4\piup \epsilon_{0}\epsilon_{r}r}
\end{equation}
with $\phi_{i\text{b}}^{*}$ being the bulk value of $\phi_{i}^{*}$. Here, $\zeta_{i}(x_1)=g_i(x_1/e_{i}=0)$ is the exclusion volume term,  $\phi_{i}=\phi_{i}(1;2)$ is the fluctuation potential at ${\bf r}_{2}$ for an ion
at ${\bf r}_{1}$, $r = r_{12}$, and $\beta =1/(k_{\text B} T)$ ($k_{\text B}$ is the Boltzmann's constant and $T$ is the absolute temperature).

	The fluctuation potential can be shown to satisfy the following set of differential equations \cite{outh3}
\begin{align}
\nabla^{2}\phi _{i}& = 0, \hspace{0.7in}   x < a/2, \label{eq4}\\                                                                                               
\nabla^{2}\phi _{i}& = -\frac{\rd^{2}\phi _{i}}{\rd x^{2}}\,, \hspace{0.38in}     r < a, \label{eq5}\\
\nabla^{2}\phi _{i}& = \kappa ^{2}(x)\phi _{i}\,, \hspace{0.325in}        x > a/2,~~  r > a, \label{eq6}
\end{align}
where  $x = x_{2}$ and $\kappa (x)$ is the local Debye-H\"{u}ckel parameter defined by
\begin{equation}
\kappa ^{2}(x)=\frac{\beta }{\epsilon _{0}\epsilon _{r}}\sum_{s}e_{s}^{2}n_{s}g_{s}(x).
\end{equation}
Equations~(\ref{eq4}) and (\ref{eq5}) are exact, while equation (\ref{eq6}) involves two approximations.
These are Loeb's closure \cite{loeb}, which gives a closed system of equations for $\phi_i$, and the linearization of the equation involving Loeb's closure. The boundary conditions associated with equations~(\ref{eq4})--(\ref{eq6}) are $\phi_i$ and $\partial\phi_{i}/\partial n$ continuous ($n$ denotes normal) at $x = a/2$ and $r = a$, while at the electrode $\phi_i$ is continuous but $\epsilon _{r}\partial\phi_i/\partial x^{+}=\epsilon _{\text w} \partial \phi_i/\partial x^{-}$. Approximate solutions of equations~(\ref{eq4})--(\ref{eq6}) have been considered for both point ions \cite{outh5} and finite size ions \cite{outh3}. The point ion solution suggests an appropriate solution in the regions $x < 0$, $0 < x < a/2$. From the analysis of reference \cite{outh5}, the limiting case of having no distance of closest approach and $k \rightarrow  0$ gives
\begin{align}
\phi_{i}& = \frac{e_i}{4\piup \epsilon_{0}\epsilon _{r}r}(1+f),      \hspace{0.38in}                  x < 0,\\
\phi_{i}& = \frac{e_i}{4\piup \epsilon_{0}\epsilon _{r}}\left(\frac{1}{r}+\frac{f}{r^{*}}\right), \qquad x>0,                                                                     
\end{align}
where $f=(\epsilon _{r}-\epsilon _{\text w})/(\epsilon _{r}+\epsilon _{\text w})$ and $r^{*}$ is the distance of the image of $e_{i}$ in the electrode from the field point. These two results suggest that an appropriate approximation for the ion size result for the equation~(\ref{eq4}) is
\begin{align}
\phi_{i}& = C\left(\frac{1}{r}+\frac{f}{r}\right), \qquad   x < 0, \label{eq10}\\
\phi_{i}& = C\left(\frac{1}{r}+\frac{f}{r^{*}}\right), \qquad   0 < x < a/2,  \label{eq11}                                                                 \end{align}
which were used in reference \cite{outh3}.

Previous work \cite{outh1,outh3,outh6,bhuiyan1,levine1,levine2} has shown that a good
approximation to the solution of equation~(\ref{eq6}) is
\begin{equation}
\phi_{i} = B\frac{\exp(-\kappa r)}{r}+B^{*}\frac{\exp(-\kappa r^{*})}{r^{*}}\,, \qquad  x > a/2, ~ r > a, \label{eq12}                             
\end{equation}
where $\kappa = \kappa (x_{1})$. The point ion result, with a distance of closest approach, indicates a complex situation such that $B$, $B^*$, $C$ can contain the imaging factor $f$.

The use of the approximate solutions~(\ref{eq10})--(\ref{eq12}) mean that the boundary conditions cannot be satisfied. An approximate fitting of the boundary conditions was considered in \cite{outh3}, giving the MPB5 theory. Here, we adopt the approach of reference \cite{outh1}. The application of Green's second identity to $\phi_{i}^{*}$ and $1/r$ in the exclusion sphere $V$, where $V$ is a truncated sphere if $a/2 < x < 3a/2$, gives, on using equation~(\ref{eq5})
\begin{align}
4\piup \lim _{r\rightarrow 0}\phi _{i}^{*}& = \int_{V}\left(\frac{1}{r}-\frac{1}{a}\right)\nabla ^{2}\psi \rd V+\frac{1}{a^{2}}\int_{S_{\text I}}\phi _{i}\rd S-\frac{1}{a}\int_{S_{\text W}}\frac{\partial \phi _{i}}{\partial n}\rd S \nonumber \\
& \quad  +\int_{S_{\text W}}\left[\frac{1}{r}\frac{\partial \phi _{i}}{\partial n}-\phi _{i}\frac{\partial}{\partial n}\left(\frac{1}{r}\right)\right]\rd S-\frac{e_{i}}{a\epsilon _{0}\epsilon _{r}}\,. \label{eq13}
\end{align}
Here, $S_{\text W}$ and $S_{\text I}$ are respectively the wall and double layer surface portions of the truncated sphere. The integrals over $S_{\text W}$ vanish for $x > 3a/2$, and $S_{\text I}$ is then the complete spherical surface. For reference, the values of the integrals appearing in equation~(\ref{eq13}), using the approximations for $\phi _{i}$, are given in the appendix.

	To determine the relations between $B$, $B^*$ and $C$ we use Gauss's theorem for ion $i$
\begin{equation}
\int_{S}\frac{\partial \phi _{i}}{\partial n}\rd S = -\frac{e_{i}}{\epsilon _{0}\epsilon _{r}}-\int_{V}\nabla ^{2}\psi \rd V. \label{eq14}
\end{equation}

Thus, combining the expressions resulting from the evaluation of equations~(\ref{eq13}), (\ref{eq14}) with equation~(\ref{eq2}) gives
\begin{align}
g_{i}(x)  &= \zeta _{i}(x)\exp\Bigg\{-\frac{1}{2}\beta e_{i}\Bigg[\left(\frac{e_{i}}{4\piup \epsilon _{0}\epsilon _{r}}\right)(F-F_{0})+F\psi (x+a)+F\psi (x-a)\nonumber\\
&\quad-\frac{(F-1)}{a}\int_{x-a}^{x+a}\psi (X)\rd X\Bigg]\Bigg\}, \label{eq15}
\end{align}
where for $a/2 < x < 3a/2$
\begin{align}
F &= \frac{A}{D}\,, \label{eq16}\\
A &= (2x + a)\re^{-y} + \frac{C}{B}(3a - 2x) + \frac{C}{B}\frac{af}{x}(a - 2x + s)
+ \frac{a}{\kappa x}\frac{B^{*}}{B}\left[\re^{-\kappa s} - \re^{-\kappa(2x + a)} \right], \label{eq17}\\
D &= (1 + y)(2x + a)\re^{-y} + \frac{C}{B}(3a - 2x) + \frac{C}{B}f(s - 2a) - \frac{a}{\kappa x}\frac{B^{*}}{B}\Big[(1 + y)\re^{-\kappa (2x + a)}\nonumber \\
  &\quad - \frac{1}{a}(a + \kappa sa - \kappa sx)\re^{-\kappa s}\Big] \label{eq18}
\end{align}                                                                                                                     and for $x > 3a/2$
\begin{equation}
F = \frac{1 + \frac{1}{2\kappa x}\left(\frac{B^{*}}{B}\right)\re^{-\kappa (2x - a)}\sinh y}{1 + y
-\frac{1}{2\kappa x}\left(\frac{B^{*}}{B}\right)\re^{-\kappa (2x - a)}(y\cosh y - \sinh y)}\,, \label{eq19}
\end{equation}
where $F_{0}=\lim_{x\rightarrow \infty}F$, $y = \kappa a$ and $s = \sqrt{a^{2} + 2xa}$.

At $x = 3a/2$, we have that $F$ is continuous but $\partial F/\partial x$ is discontinuous in general. The expression for $F$ depends upon the ratios $C/B$, $B^{*}/B$ for $a/2 < x < 3a/2$ and on the single ratio $B^{*}/B$ for $x > 3a/2$. Putting $C = B\exp(-y)$, $B^{*} = Bf\exp(-y+\kappa s)$ in $a/2 < x < 3a/2$ and $B^{*} = Bf\exp y$ in $x > 3a/2$ gives the MPB5 theory \cite{outh3}. Early calculations of $F$ did not consider the exclusion region $0 < x < a/2$ for $\phi _{i}$ so that $F$ is defined in the two regions $a/2 < x < a$ and $x > a$, for example the MPB4 theory \cite{outh1,outh2}. In these earlier works, the value of $F$ differs from that of the MPB5 when $x > 3a/2$. On putting $B^{*} = fB$ in equation~(\ref{eq19}), the $F$ of MPB4 for $x > a$ is derived. This suggests an alternative $F$, analogous to that for MPB4, which is given by $C = B$ in equations~(\ref{eq17}), (\ref{eq18}) and $B^{*} = fB$ in equations~(\ref{eq17})--(\ref{eq19}). There are many possibilities of choosing the ratios $C/B$, $B^{*}/B$, depending upon the chosen criterion. However, a better approach may be to improve upon the analysis of equations~(\ref{eq4}), (\ref{eq5}) and the nonlinear version of equation~(\ref{eq6}).

Various expressions have been used for evaluating the exclusion volume term $\zeta_{i}(x)$ in equation~(\ref{eq2})
\cite{outh1,outh2,outh7}. The most accurate exclusion volume terms are those based on the BBGY expansion \cite{outh2} or charging up the wall and introducing the non-uniform direct correlation function \cite{outh7}. Here, we will use the BBGY approach, where for $x > a/2$
\begin{equation}
\zeta _{i}(x)=\exp\left\{-2\piup \int_{\infty}^{x}\sum_{s}n_{s}\zeta _{is}(a)\int_{\text{max}(\frac{a}{2},y-a)}^{y+a}
(X-y)g_{s}(X)\exp[-\beta e_{i}\Phi (y,X)]\rd X\rd y\right\}
\end{equation}
with $\zeta_{i}(x)=0$ for $x < a/2$.  The contact value $\zeta _{ij}(a)$  between ions $i$ and $j$ is that proposed by Fischer \cite{fischer} while $\Phi(y,X)=\phi (1;2/e_{s}=0/r_{12}=a)$.

\subsection{Monte Carlo simulations}

The Monte Carlo simulations were performed in the canonical ($N, V, T$) ensemble using the standard and widely used Metropolis algorithm \cite{metropolis,allen}. The techniques were the same as those used in some of our earlier works (see for example, references \cite{bhuiyan2,bhuiyan3}). The MC cell was a rectangular parallelepiped of dimensions $L_{x}$, $L_{y}$, and $L_{z}$, with $L_{y} = L_{z}$. One of the $y$-$z$ faces was a charged wall, viz., the electrode with a uniformly distributed surface charge at $x = 0$, while the other at $x = L_{x}$ was a neutral wall. The surface charge density on the electrode was determined from
\begin{equation}
\sigma = -\frac{(N_{+}|Z_{+}|-N_{-}|Z_{-}|)e}{L_{y}^{2}}\,,
\end{equation}		
where $N_{+}$ and $N_{-}$ are the number of cations and anions, respectively. It is to be noted that with the above definition of $\sigma $, the local electro-neutrality at the wall is built in. In order to simulate the semi-infinite system, conventional minimum image techniques along with periodic boundary conditions in the $y$ and $z$ directions were implemented \cite{allen}. The effect of the long range nature of the Coulomb interactions was treated by the parallel charge sheets method pioneered by Torrie and Valleau \cite{torrie1}. This procedure was later examined and improved upon by Boda et al. \cite{boda}, who also confirmed the validity of the method for charge hard spheres.

The situation when the permittivity of the medium of the charged wall differs from that of the electrolyte solution has also been treated by Torrie at al. \cite{torrie3} and their procedure adopted here. These authors regarded the polarizability of the electrode as a classic electrostatic image problem, viz., the total surface charge density $\sigma $ has a contribution from the fictitious image charge $-f\sigma $  due to the polarization of the electrode. In passing we would like to note that the Torrie and co-workers did their simulations in the grand canonical ($\mu, V, T$) ensemble unlike the canonical ensemble in the present case.

In the canonical ensemble, the bulk target concentration can only be achieved by a trial-and-error method of adjusting the cell length $L_{x}$ and/or the number of particles. We used both of these methods where necessary and imposed a tolerance of less than 2$\%$ error in reproducing the bulk concentration.  Typically, around 400 particles were simulated and approximately $10^{8}$ configurations sampled out of which the first $10^{7}$ (10$\%$ of the total number of configurations) were used for system equilibration before statistics were taken.

\section{Results}

The properties of the diffuse double layer are found by solving the Poisson equation,
equations~(\ref{eq1}) and (\ref{eq15}), for the mean electrostatic potential and hence from equation~(\ref{eq15}) the singlet distribution function. Three different values of $F$ were treated in detail,
these corresponding to those of MPB4, MPB5, and the new $F$ given by $C = B$, $B^{*} = fB$ in equations~(\ref{eq16})--(\ref{eq19}), respectively. A quasi-linearisation technique was used \cite{outh1,bellman} which has been
successfully employed in past investigations. The parameters chosen were those used by
Wang \cite{wang3} who simulated charge asymmetric 1:3 and 3:1 RPM electrolytes with imaging.
For the RPM
electrolyte, we treat 1:2, 2:1 and some 1:3, 3:1 valences with $a =4\cdot 10^{-10}$~m,
$\epsilon _{r}= 80$, $T =298$~K while the wall is characterized by $\epsilon _{\text w} = 2, 80$ or $\infty $.
When $\epsilon_{\text w}\neq \epsilon _{r}$, imaging occurs which is repulsive for $\epsilon_{\text w} < \epsilon _{r}$
and attractive for $\epsilon_{\text w} > \epsilon _{r}$. The value $\epsilon _{\text w} =$ 2, which is relevant
for a medium such as silica, while $\epsilon _{\text w} =\infty $ represents a metallic electrode.

\begin{figure}[!t]
\centering
\begin{minipage}{0.495\textwidth}
\begin{center}
\includegraphics[width=1.00\textwidth]{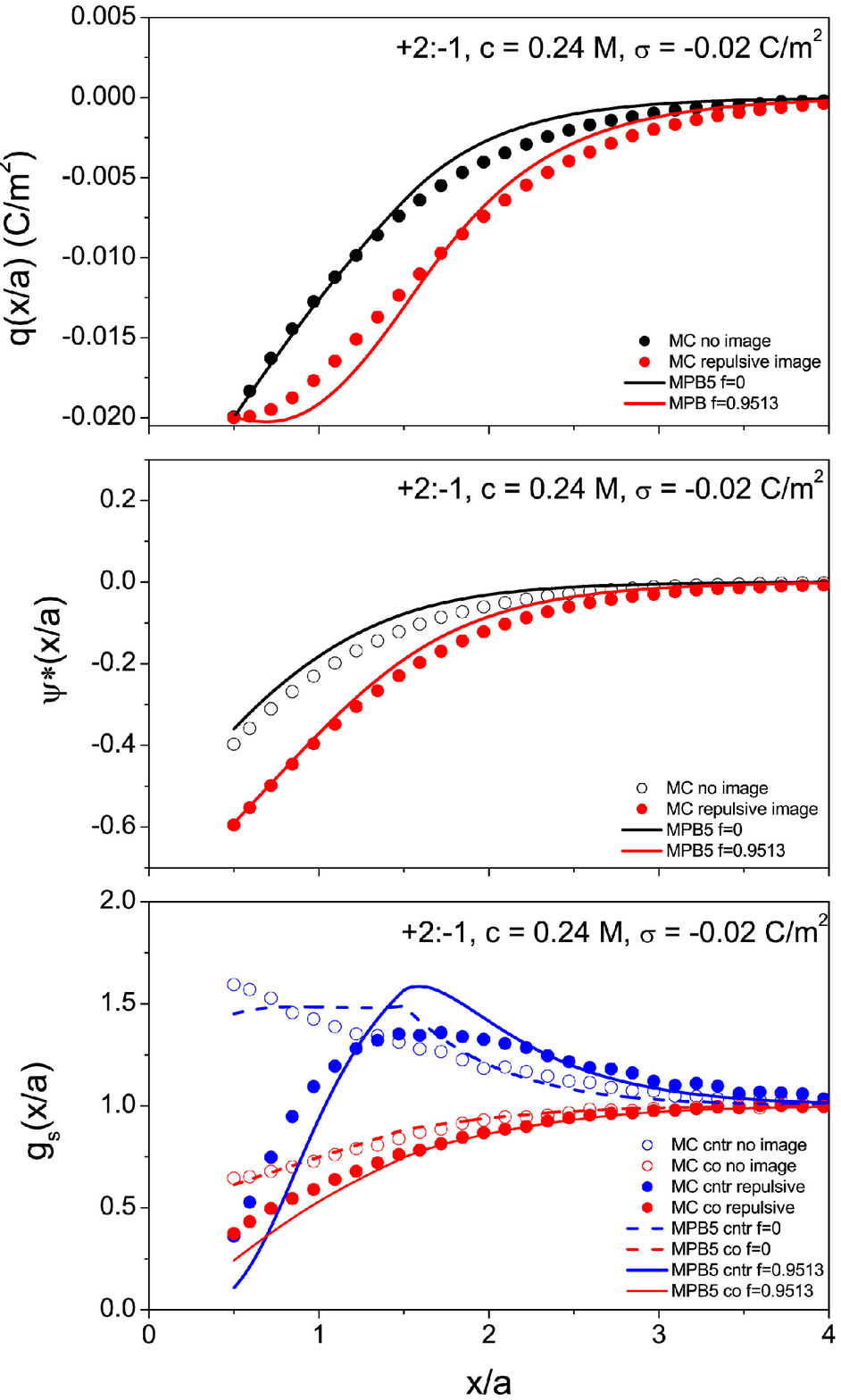}
\end{center}
\end{minipage}
\begin{minipage}{0.495\textwidth}
\begin{center}
\includegraphics[width=1.00\textwidth]{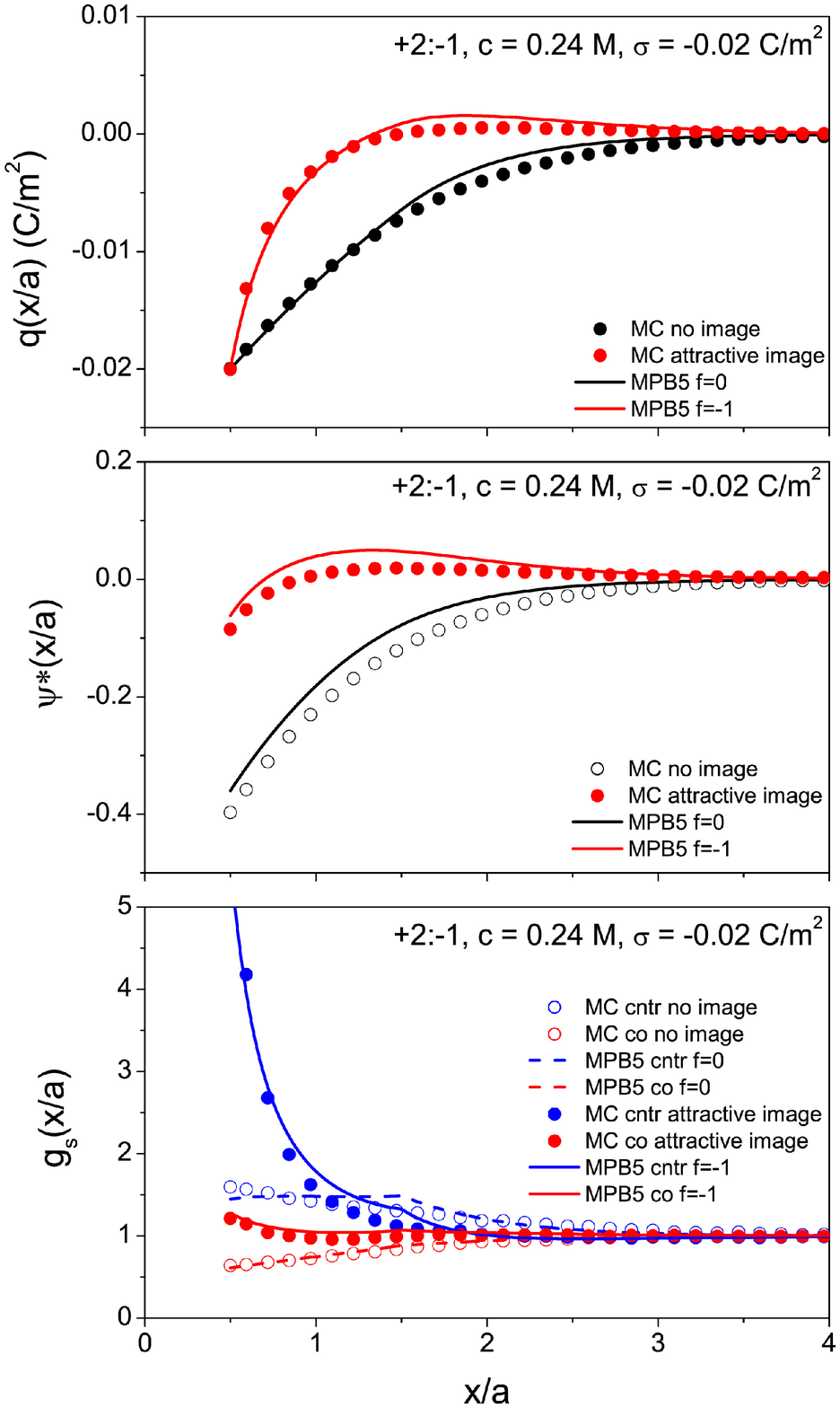}
\end{center}
\end{minipage}
\begin{minipage}{0.495\textwidth}
\caption{(Color online) The MPB5 electrode-ion singlet distribution functions
$g_{s}(x/a)$ (lower panel), the mean electrostatic potential $\psi ^{*}(x/a)$
(middle panel), and the integrated charge $q(x/a)$ (upper panel) as functions
of $x/a$ for a 2:1 electrolyte. The symbols represent MC data, while the lines
represent the MPB results. The notations ``co'' and ``cntr'' stand for ``coion'' and
``counterion'', respectively. Legend as given in the figure.}
\label{fig1}
\end{minipage}
\begin{minipage}{0.495\textwidth}
\vspace{-11.5mm}
\caption{(Color online) The MPB5 electrode-ion singlet distribution functions
$g_{s}(x/a)$ (lower panel), the mean electrostatic potential $\psi ^{*}(x/a)$
(middle panel), and the integrated charge $q(x/a)$ (upper panel) as functions
of $x/a$ for a 2:1 electrolyte. Notation as in figure~\ref{fig1} and legend as given
in the figure.}
\label{fig2}
\end{minipage}
\end{figure}

We start by looking in detail at the predictions of the MPB5 theory for 2:1 electrolytes
with and without imaging. Figures~\ref{fig1} and \ref{fig2} treat the repulsive ($f = 0.9513$) and attractive
($f = -1$) cases, respectively, at electrolyte concentration $c = 0.24$~mol/dm$^{3}$ and
surface charge density $\sigma = -0.02$~C/m$^{2}$. Displayed are the singlet distribution
functions $g_{s}(x/a)$, the dimensionless mean electrostatic potential
$\psi ^{*}(x/a)=|e|\beta \psi (x/a)$ and the integrated charge density $q(x/a)$ defined by
\begin{equation}
q(x)=\sigma + \sum _{s}n_{s}e_{s}\int_{0}^{x}g_{s}(y)\rd y.                                                                         
\end{equation}
The integrated charge gives a measure of the total charge within a distance $x$
of the electrolyte and enables a succinct illustration of the two phenomena CR and SCA.

For no imaging, the simulated singlet distribution functions apparently display a
monotonous behaviour of the co and counter ions, with a corresponding monotonous behaviour of
the mean electrostatic potential and the integrated charge. The MPB5 theory has a small shoulder
in the counterion $g_{s}(x)$ at $x \sim 3a/2$ arising from the discontinuity of $\partial F/\partial x$
at $x = 3a/2$. This is an unfortunate feature at low concentrations which has been observed
earlier \cite{outh3}. The influence of this feature is not observed in $\psi ^{*}(x)$ since $\psi ^{*}(x)$
and its derivative are continuous everywhere. The principal effect of imaging in the distribution
functions for the repulsive situation is that the divalent counterion is attracted for large $x$
but reduces and becomes small near the wall,
while a slight reduction is seen in the monovalent coion.
Conversely for the attractive situation, the coion profile increases to a value greater than 1
near the electrode, and the counterion profile rapidly increases in the neighborhood of the electrode.
These imaging effects are reflected in the reduction (repulsive case) and increase
(attractive case) over the no imaging situation, of the mean electrostatic potential
and integrated charge. A very small CR is seen when $f = -1$. The MPB5 accurately predicts
$\psi ^{*}(x)$ and $q(x)$, and apart from the shoulder feature, the
$g_{s}(x)$ are quantitatively or qualitatively correct.

\begin{figure}[!t]
\centering
\begin{minipage}{0.495\textwidth}
\begin{center}
\includegraphics[width=1.00\textwidth]{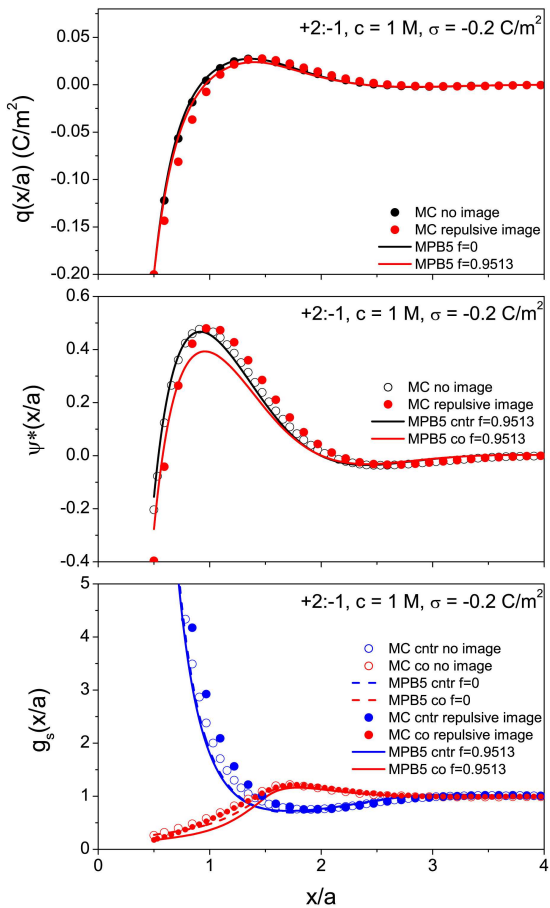}
\end{center}
\end{minipage}
\begin{minipage}{0.495\textwidth}
\begin{center}
\includegraphics[width=1.00\textwidth]{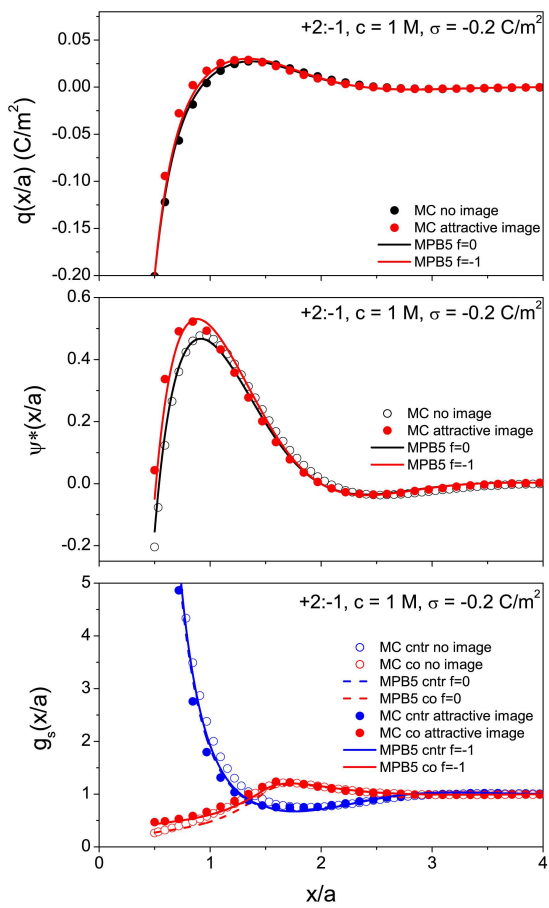}
\end{center}
\end{minipage}
\begin{minipage}{0.495\textwidth}
\caption{(Color online) The MPB5 electrode-ion singlet distribution functions
$g_{s}(x/a)$ (lower panel), the mean electrostatic potential $\psi ^{*}(x/a)$
(middle panel), and the integrated charge $q(x/a)$ (upper panel) as functions
of $x/a$ for a 2:1 electrolyte. Notation as in figure~\ref{fig1} and legend as given
in the figure.}
\label{fig3}
\end{minipage}
\begin{minipage}{0.495\textwidth}
\caption{(Color online) The MPB5 electrode-ion singlet distribution functions
$g_{s}(x/a)$ (lower panel), the mean electrostatic potential $\psi ^{*}(x/a)$
(middle panel), and the integrated charge $q(x/a)$ (upper panel) as functions
of $x/a$ for a 2:1 electrolyte. Notation as in figure~\ref{fig1} and legend as given in
the figure.}
\label{fig4}
\end{minipage}
\vspace{3mm}
\end{figure}

\begin{figure}[!t]
\centering
\begin{minipage}{0.495\textwidth}
\begin{center}
\includegraphics[width=1.00\textwidth]{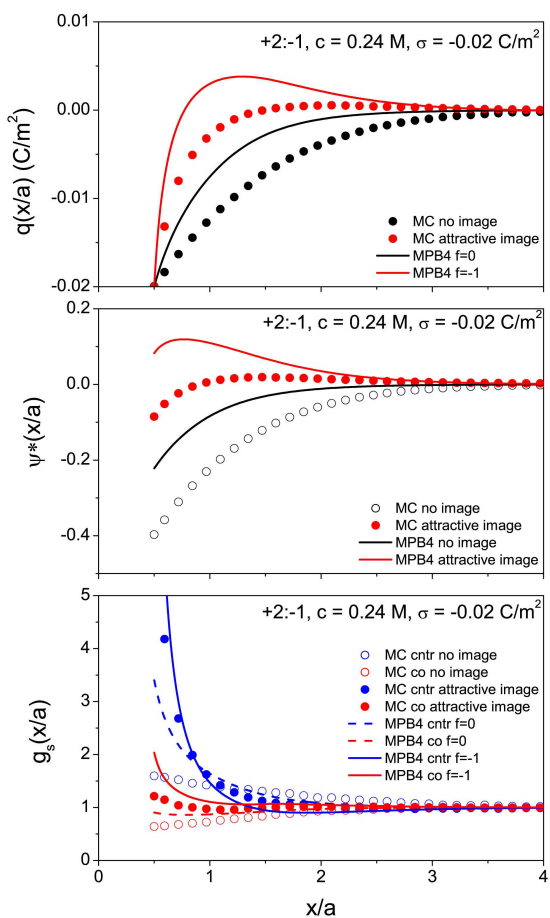}
\end{center}
\end{minipage}
\begin{minipage}{0.495\textwidth}
\begin{center}
\includegraphics[width=1.00\textwidth]{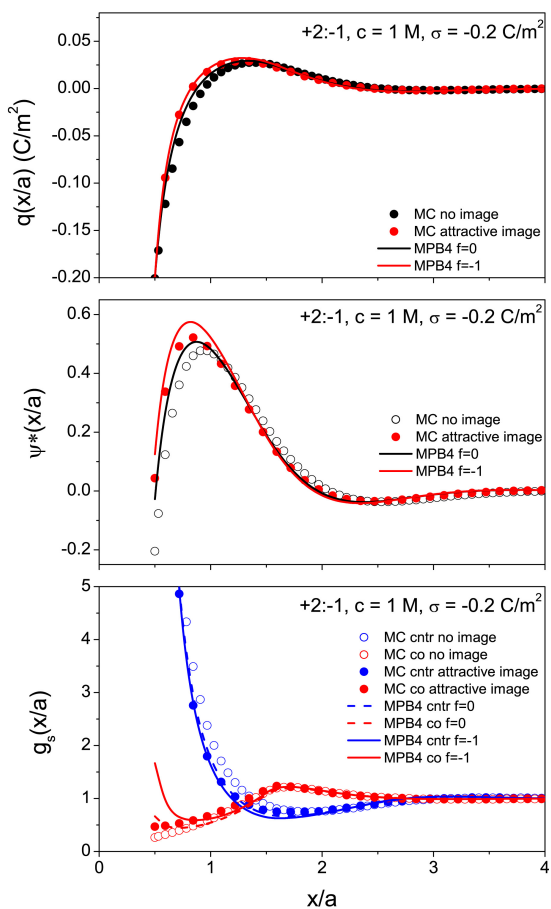}
\end{center}
\end{minipage}
\begin{minipage}{0.495\textwidth}
\caption{(Color online) The MPB4 electrode-ion singlet distribution functions
$g_{s}(x/a)$ (lower panel), the mean electrostatic potential $\psi ^{*}(x/a)$
(middle panel), and the integrated charge $q(x/a)$ (upper panel) as functions
of $x/a$ for a 2:1 electrolyte. Notation as in figure~\ref{fig1} and legend as given in
the figure.}
\label{fig5}
\end{minipage}
\begin{minipage}{0.495\textwidth}
\caption{(Color online) The MPB4 electrode-ion singlet distribution functions
$g_{s}(x/a)$ (lower panel), the mean electrostatic potential $\psi ^{*}(x/a)$
(middle panel), and the integrated charge $q(x/a)$ (upper panel) as functions
of $x/a$ for a 2:1 electrolyte. Notation as in figure~\ref{fig1} and legend as given
in the figure.}
\label{fig6}
\end{minipage}
\vspace{3mm}
\end{figure}

\begin{figure}[!t]
\centering
\begin{minipage}{0.495\textwidth}
\begin{center}
\includegraphics[width=1.00\textwidth]{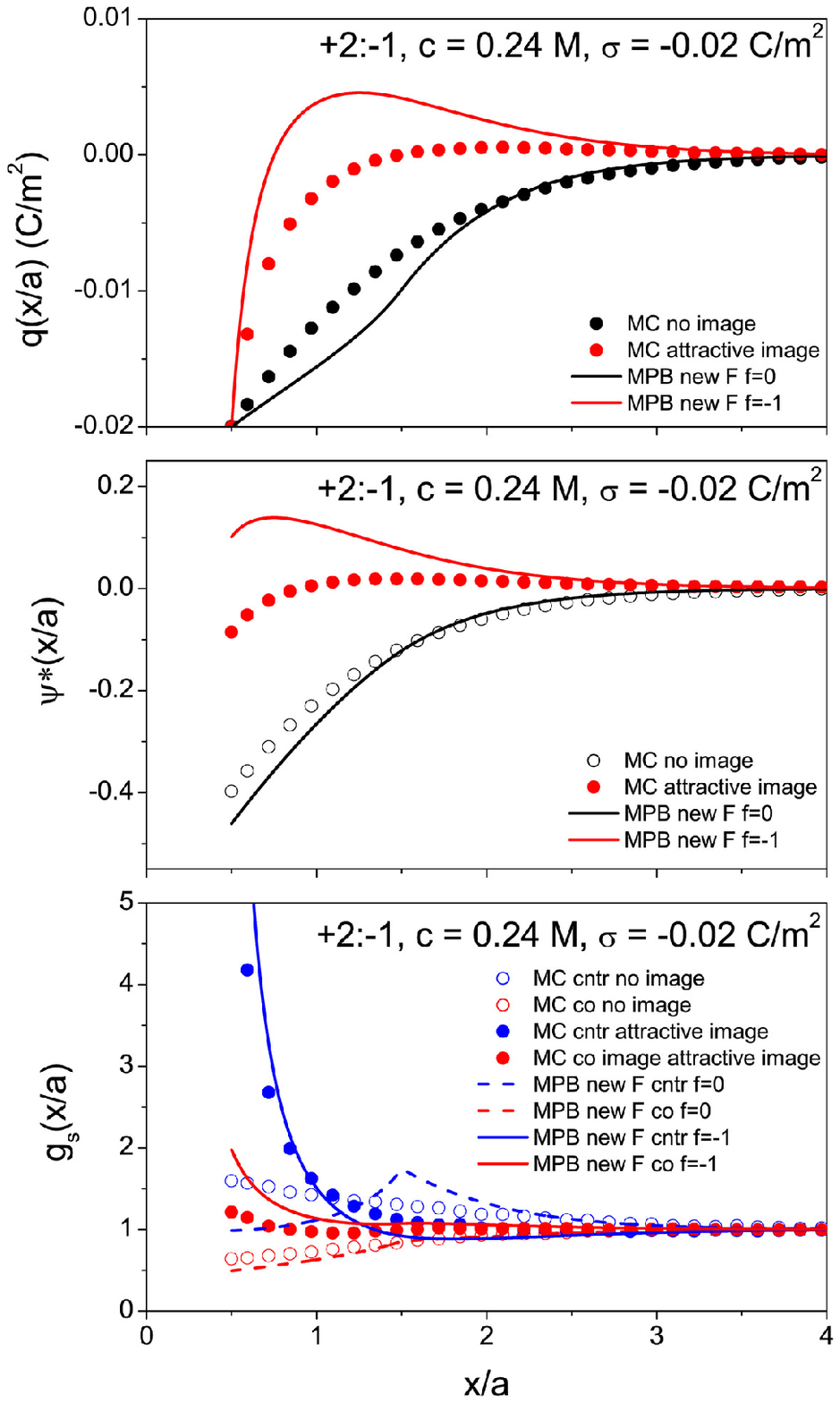}
\end{center}
\end{minipage}
\begin{minipage}{0.495\textwidth}
\begin{center}
\includegraphics[width=1.00\textwidth]{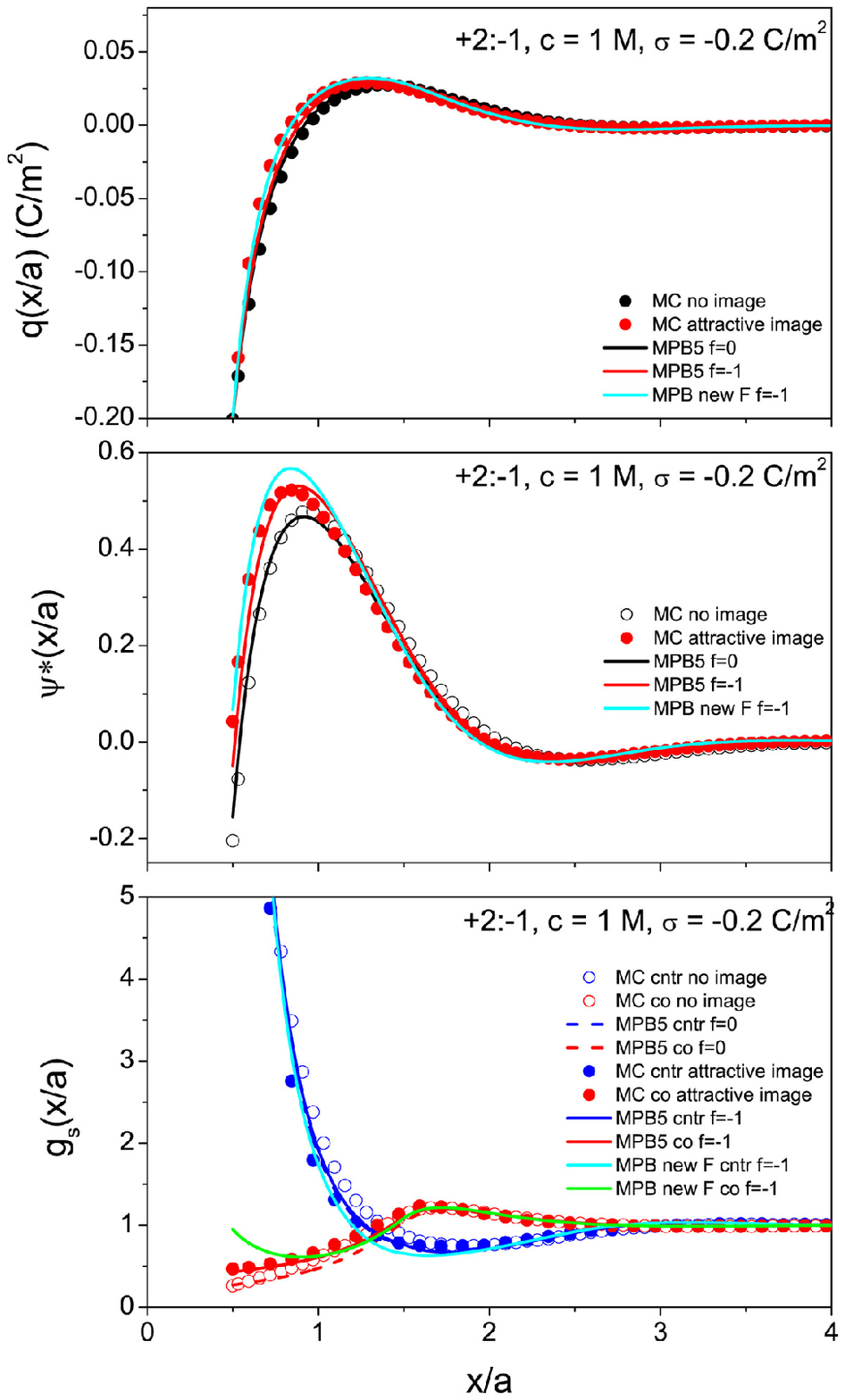}
\end{center}
\end{minipage}
\begin{minipage}{0.495\textwidth}
\caption{(Color online) The MPB (new $F$) electrode-ion singlet distribution functions
$g_{s}(x/a)$ (lower panel), the mean electrostatic potential $\psi ^{*}(x/a)$
(middle panel), and the integrated charge $q(x/a)$ (upper panel) as functions
of $x/a$ for a 2:1 electrolyte. Notation as in figure~\ref{fig1} and legend as given
in the figure.}
\label{fig7}
\end{minipage}
\begin{minipage}{0.495\textwidth}
\caption{(Color online) The MPB5 and MPB (new $F$) electrode-ion singlet distribution functions
$g_{s}(x/a)$ (lower panel), the mean electrostatic potential $\psi ^{*}(x/a)$
(middle panel), and the integrated charge $q(x/a)$ (upper panel) as functions
of $x/a$ for a 2:1 electrolyte. Notation as in figure~\ref{fig1} and legend as given
in the figure.}
\label{fig8}
\end{minipage}
\end{figure}

At the higher concentration $c = 1$~mol/dm$^{3}$ and surface charge $\sigma = -0.2$~C/m$^{2}$,
there is a clear change in the electrolyte properties, viz., figures~\ref{fig3}, \ref{fig4}. The striking feature
is the obvious damped oscillatory behaviour of various functions for all three situations
($f = 0.9513$, $f = 0$, $f = -1$). The onset of this oscillatory behaviour mainly depends,
at a small surface charge, upon the value of $y_{0}=\kappa _{0}a$, where $\kappa _{0}$ is the
bulk Debye-H\"{u}ckel constant. A damped oscillatory behaviour in $\psi ^{*}(x)$ is predicted
in the linear MPB theory for $y_{0} >$ 1.2412 \cite{outh8a} for symmetric valences
and for $y_{0} >$ 1.15 \cite{bhuiyan1} for 2:1/1:2 valences. However, the latter is probably
too high since in simulation and the HNC, the $g_{s}(x)$ value for 1:2 homogeneous electrolytes
implies $y_{0} \approx  0.76$ \cite{ulander}. Here, $y_{0} = 2.26$, while the previous
concentration of 0.24~mol/dm$^{3}$ gave
$y_{0} = 1.11$, so that a qualitative change in the behaviour between the two concentrations
is predicted by the MPB theories. With no imaging, the initial maximum in $\psi ^{*}(x)$ is
at $x \sim a$ with CR starting at about the same value of $x$ and having a maximum at
$x \sim  5a/4$. Both the repulsive and attractive cases are qualitatively similar
to that for no imaging, showing small but different deviations from $f = 0$. For example,
with the mean potential, the first maximum is shifted to a larger
$x$ for the repulsive case,
but closer to the electrode with a larger maximum for the attractive case. At this higher
concentration, there is no shoulder in the MPB5 $g_{s}(x)$ at $x \sim 3a/2$, and the MPB5
gives an accurate picture of all three functions. Clearly, the effect of
discontinuity in $\partial F/\partial x$ at $x = 3a/2$ gets masked at high concentrations.
The MPB4 theory seems a possible alternative to the MPB5 theory as it does not have
this discontinuity
at $x = 3a/2$. Figures~\ref{fig5} and \ref{fig6} show the MPB4 for the 2:1 attractive cases. Clearly, the
theory is inadequate, especially for the lower concentration. The main inaccuracy is the
unphysical upturn in the coion distribution function close to the wall. A further possible
alternative theory is that given by $C = B$, $B* = fB$ in equations~(\ref{eq16})--(\ref{eq19}).
Looking again at the 2:1 attractive cases, figures~\ref{fig7} and \ref{fig8}, this theory incorrectly
predicts (i) an unphysical rise in the coion $g_{s}(x)$ near the wall, (ii) a shoulder
in the counterion $g_{s}(x)$ for no imaging at the lower concentration. Apart from the
unphysical behaviour (i), the theory based on the new $F$ gives a good representation of
the mean potential and integrated charge at $c = 1$~mol/dm$^{3}$. Unfortunately, neither
the MPB4 nor that based on the new $F$ is an improvement on the MPB5.

\begin{figure}[!t]
\centering
\begin{minipage}{0.495\textwidth}
\begin{center}
\includegraphics[width=0.955\textwidth]{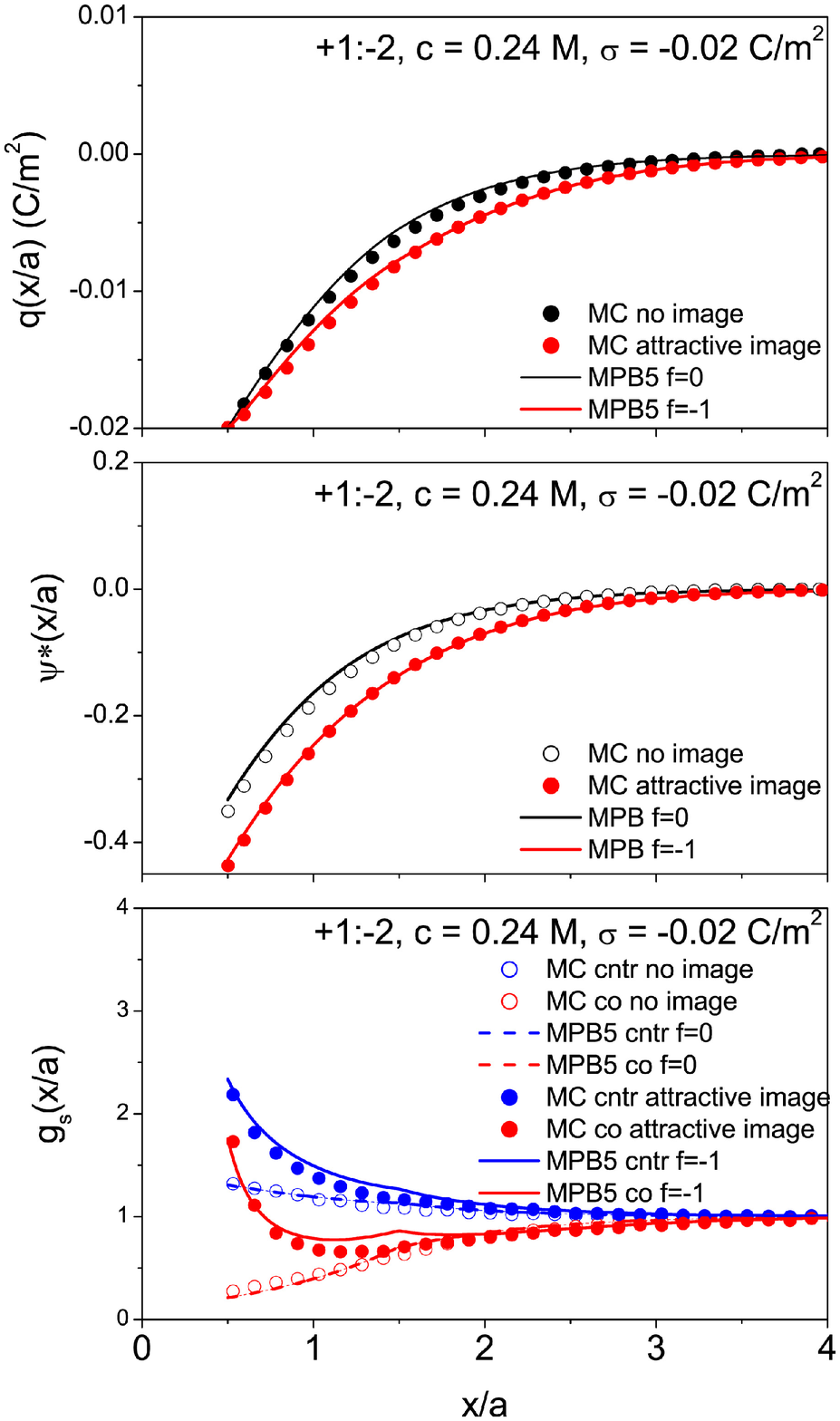}
\end{center}
\end{minipage}
\begin{minipage}{0.495\textwidth}
\begin{center}
\includegraphics[width=1.00\textwidth]{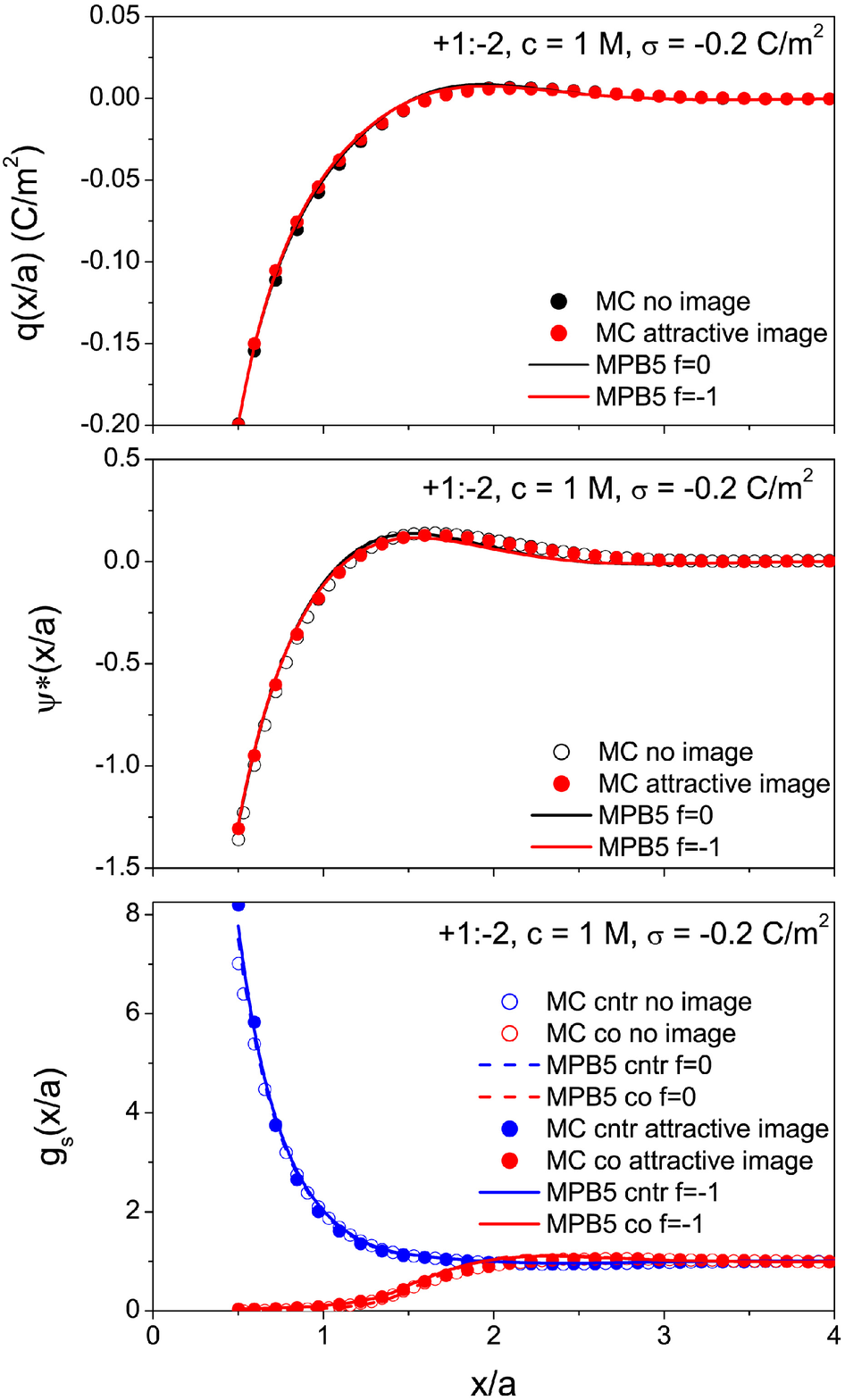}
\end{center}
\end{minipage}
\vspace{-2mm}
\begin{minipage}{0.495\textwidth}
\caption{(Color online) The MPB5 electrode-ion singlet distribution
functions $g_{s}(x/a)$ (lower panel), the mean electrostatic potential $\psi ^{*}(x/a)$
(middle panel), and the integrated charge $q(x/a)$ (upper panel) as functions
of $x/a$ for a 1:2 electrolyte. Notation as in figure~\ref{fig1} and legend as given
in the figure.}
\label{fig9}
\end{minipage}
\begin{minipage}{0.495\textwidth}
\caption{(Color online) The MPB5 electrode-ion singlet distribution functions
$g_{s}(x/a)$ (lower panel), the mean electrostatic potential $\psi ^{*}(x/a)$
(middle panel), and the integrated charge $q(x/a)$ (upper panel) as functions
of $x/a$ for a 1:2 electrolyte. Notation as in figure~\ref{fig1} and legend as given
in the figure.}
\label{fig10}
\end{minipage}
\vspace{-1mm}
\end{figure}

Figures~\ref{fig9} and \ref{fig10} consider the prediction of the MPB5 theory for the 1:2
electrolyte where now the coion is divalent. Counterion attraction with the wall plays a
dominant role in shaping the structural properties of the electric double layer and hence
many of the 1:2 double layer properties are similar to those of the 1:1 case,
while the 2:1 properties resemble that for the 2:2 situation (see for example, reference
\cite{bhuiyan4}). Thus, it comes as no surprise
that the predicted structure for the 1:2 cases are relatively closer to the simulations
than that seen earlier for the 2:1 cases. Only
attractive imaging is treated in figures~\ref{fig9}, \ref{fig10} since any deviation from no imaging tends to
be greater than those for the repulsive case. At a low concentration, figure~\ref{fig9}, the
coion is attracted near the wall, and since it is divalent, it has a greater contact value than
in the monovalent case in figure~\ref{fig2}. The difference in the valence of the counterion means that
the deviations from no imaging for the mean potential and an integrated charge are less and
oppositely directed. At a higher concentration, figure~\ref{fig10}, the imaging has little effect.
There is still a damped oscillatory feature in the functions, with CR occurring, but it is
not as pronounced as for the divalent counterion. Overall, the MPB5 theory accurately predicts
the 1:2 double layer, apart from the behaviour of the coion in the neighborhood of $x = 3a/2$.
The alternative MPB4 and new $F$ theories perform well in the 1:2 case. 

\begin{figure}[!t]
\centerline{\includegraphics[width=0.5\textwidth]{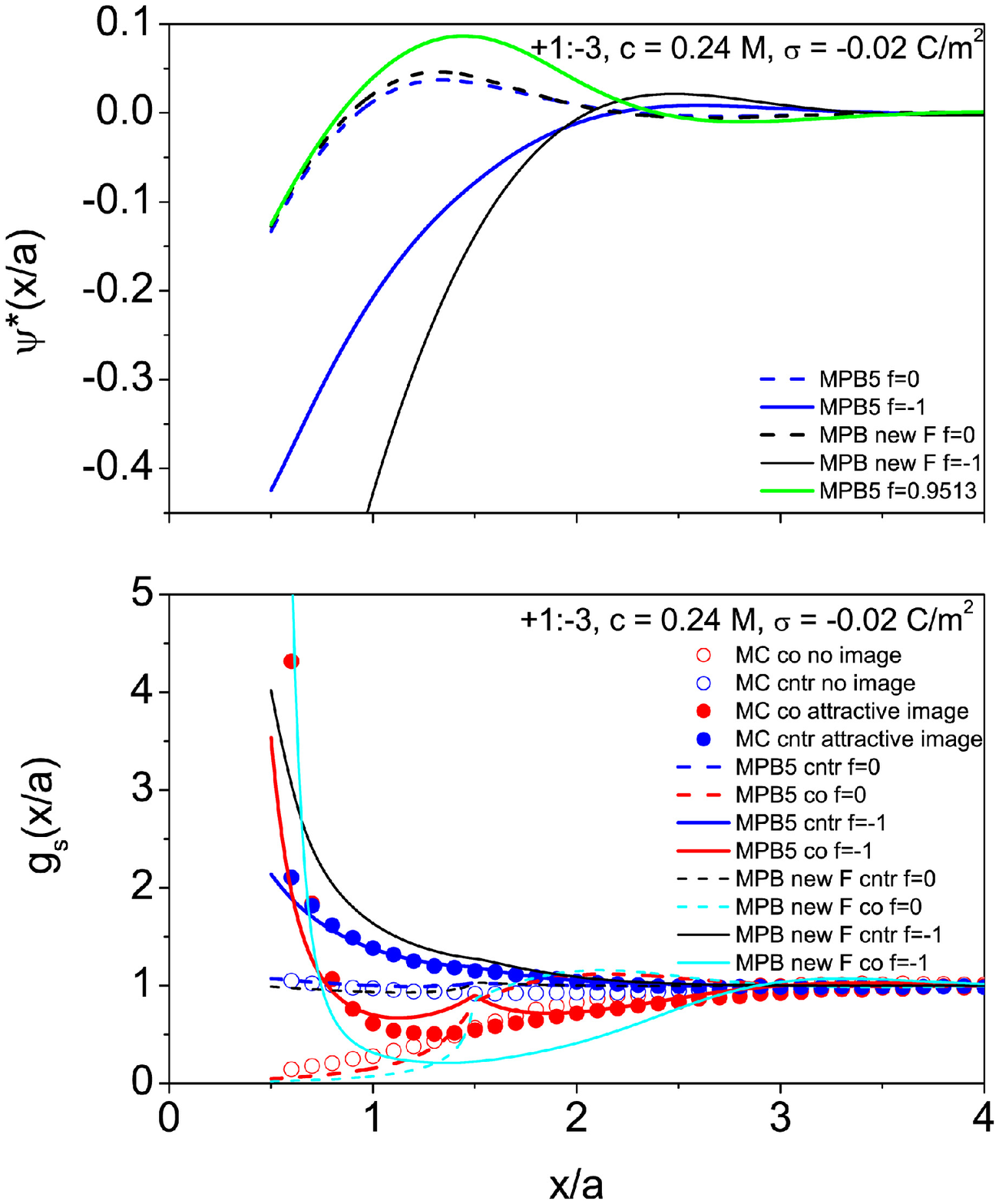}}
\caption{(Color online) The MPB5 and MPB (new $F$) electrode-ion singlet distribution functions
$g_{s}(x/a)$ (lower panel) and the mean electrostatic potential $\psi ^{*}(x/a)$
(upper panel) as functions of $x/a$ for a 1:3 electrolyte. Notation as in figure~\ref{fig1} and
legend as given in the figure. The MC data are from reference~\cite{wang3}.}
\label{fig11}
\end{figure}

\begin{figure}[!t]
\centering
\begin{minipage}{0.495\textwidth}
\begin{center}
\includegraphics[width=1.00\textwidth]{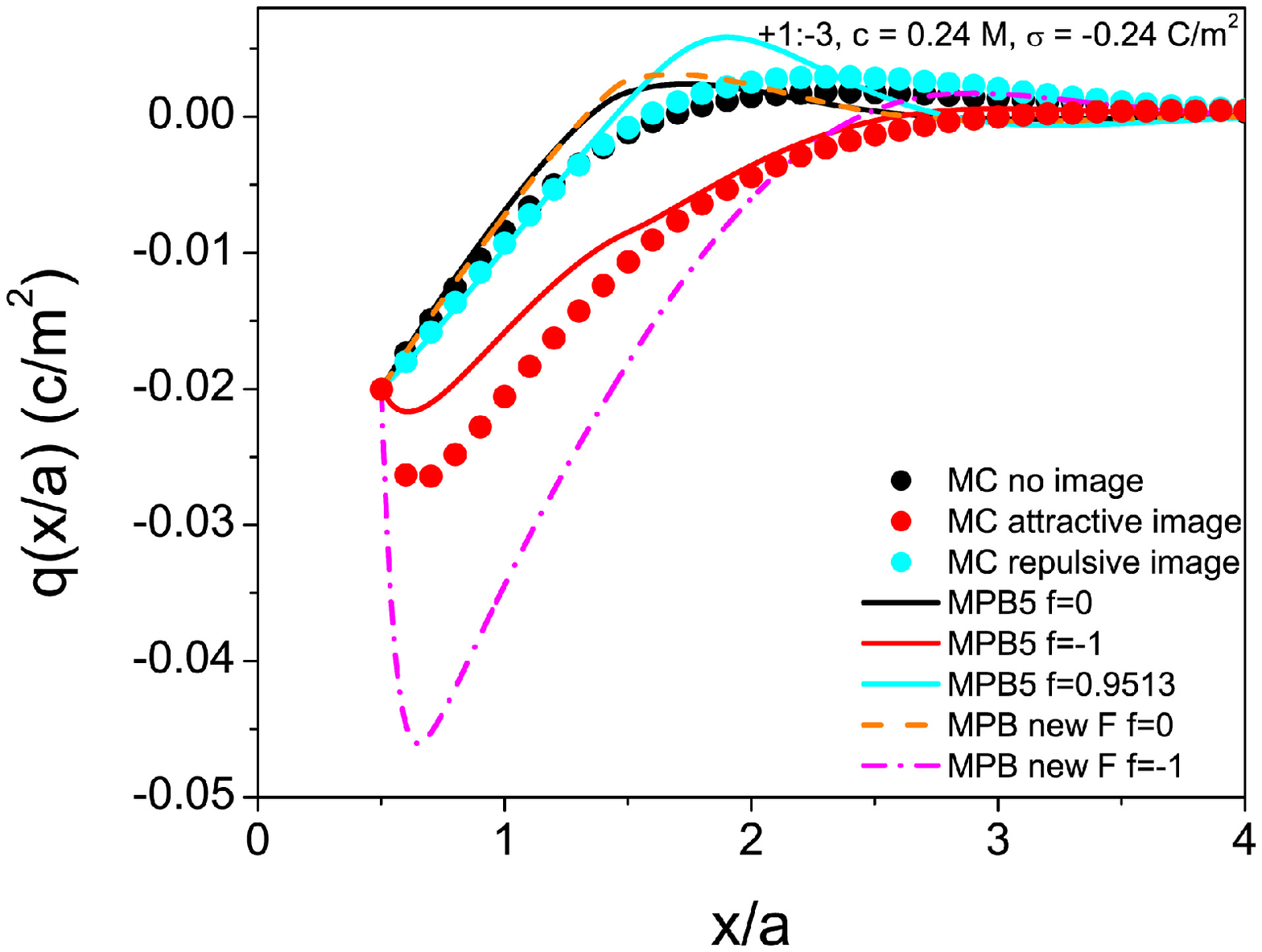}
\end{center}
\end{minipage}
\begin{minipage}{0.495\textwidth}
\begin{center}
\includegraphics[width=1.00\textwidth]{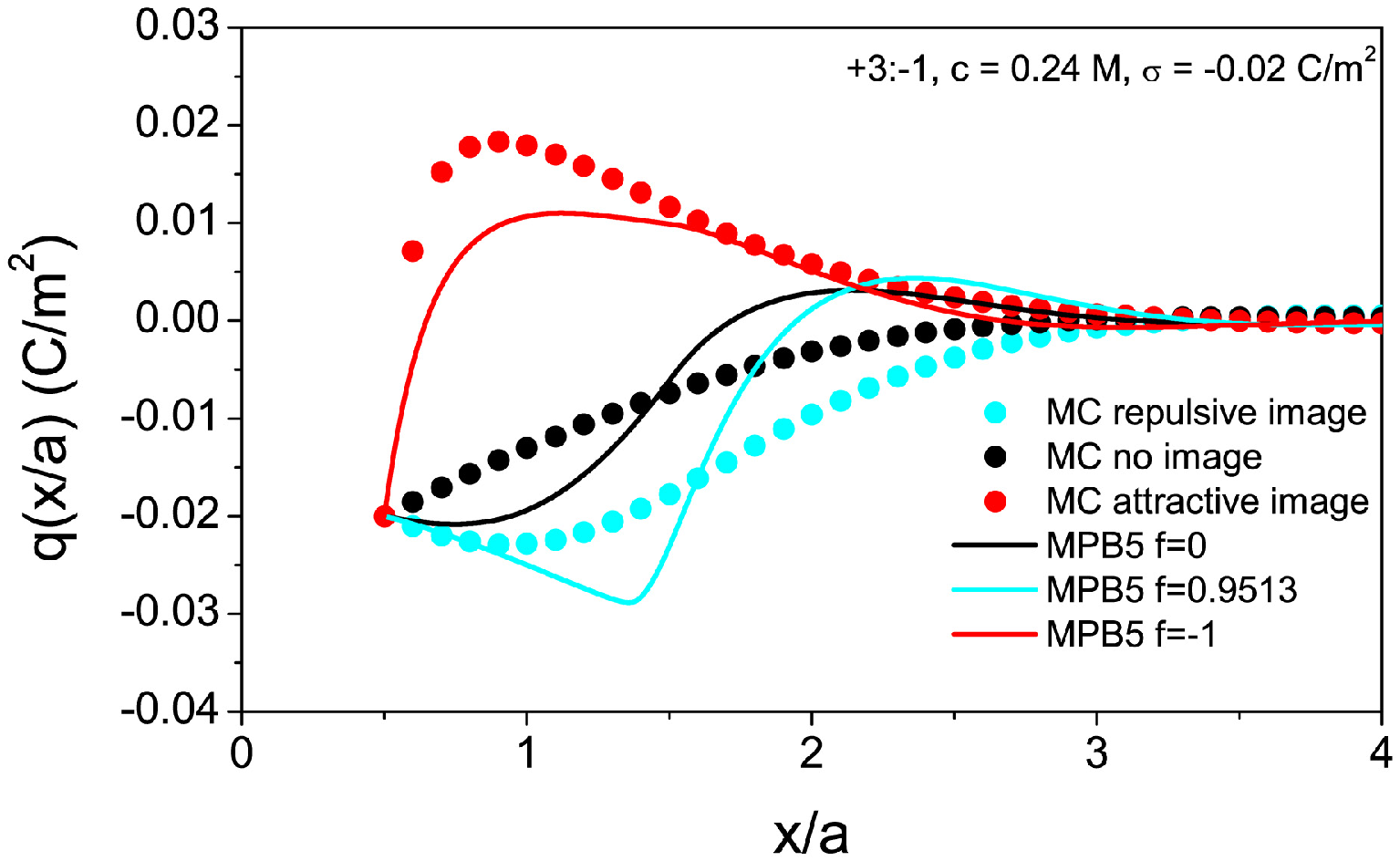}
\end{center}
\end{minipage}
\begin{minipage}{0.495\textwidth}
\caption{(Color online) The MPB5 and MPB (new $F$) integrated charge $q(x/a)$ as a function
of $x/a$ for a 1:3 electrolyte. The symbols represent MC data, while the lines represent
the MPB results. Legend as given in the figure.
The MC data are from reference~\cite{wang3}.}
\label{fig12}
\end{minipage}
\begin{minipage}{0.495\textwidth}
\vspace{-4mm}
\caption{(Color online) The MPB5 integrated charge $q(x/a)$ as a function
of $x/a$ for a 3:1 electrolyte. The symbols represent MC data,
while the lines represent the MPB results. Legend as given in the figure.
The MC data are from reference~\cite{wang3}.}
\label{fig13}
\end{minipage}
\end{figure}

Lastly, we briefly look at the MPB theory for a 1:3 and 3:1 electrolyte with the imaging at
$c = 0.24$~mol/dm$^{3}$ and $\sigma=-0.02$~C/m$^{2}$. Replacing the divalent coion by a
trivalent ion leads to CR ($f = 0.9513$, $f = 0$) and also to SCA when there is an attractive
imaging ($f = -1$), viz., figures~\ref{fig11}, \ref{fig12}. Alternatively, when the trivalent ion is the counterion
there is again CR and SCA, but now SCA occurs when there is a repulsive imaging and CR for an attractive
imaging, viz., figure~\ref{fig13}. MC simulations \cite{wang3} indicate that the MPB5 prediction of SCA
for no imaging is incorrect. The CR is expected as $y_{0} = 1.5636$, while the counterintuitive
SCA arises solely through polarization forces. In the 1:3 situation, the attraction of the coion
image overcomes the repulsion of the surface charge, while for the 3:1 electrolyte, a much
larger repulsive force acting on the counterion is sufficient to neutralise the screening in the
neighborhood of the wall. SCA does not occur for either 1:2 or 2:1 electrolytes at these parameters.
Higher valences provide a stringent test for theories. The discrepancies in $g_{s}(x)$ mean that the
MPB5 and the new $F$ theory predict both CR and SCA with varying degrees of success.

\section{Conclusions}

	In this paper, we have looked again at the formulation of a
modified Poisson-Boltzmann equation for the planar electric double layer
formed by a restricted primitive model electrolyte. The derivation of the
MPB equation follows the pattern and techniques utilized in some of our
earlier works \cite{outh1,outh2} but is a bit more general. Our focus
was on obtaining a general expression for the quantity $F$, a key ingredient
in the theory, which incorporates important aspects of inter-ionic
correlations. We have shown that an earlier version of the theory, viz.,
MPB5 \cite{outh3}, follows as a special case of the generalized $F$.
Although the MPB4 version does not follow directly, the relationship between
these two versions can be clearly assessed. With such features built in,
the theory is more flexible and it is likely that the MPB equation will
become more adaptable to particular physical situations.

    The MPB theory does not satisfy exactly the contact condition
\cite{henderson2,henderson3,martin} which holds for no imaging. Previous work
\cite{outh8,bratko,silvestre} has shown that the MPB5
theory fairly accurately satisfies the contact condition for 1:1 electrolytes  and gives a good agreement for 1:2/2:1 and 2:2 electrolytes. A
similar conclusion holds in general when the new $F$ is used, but the MPB4
theory can be poor. Adapting the MPB approach to satisfy the contact
condition would be useful for inner consistency. However, for the present
univalent and divalent RPM electrolytes, the MPB5 theory has shown to be
in good overall agreement with both structural and thermodynamic simulation
properties \cite{bhuiyan4}.

   We have studied the effect of imaging on 1:2, 2:1 valence electrolytes with the emphasis on the
phenomenon of charge reversal. Also briefly treated were 1:3, 3:1 electrolytes where, besides the
charge inversion, polarization can produce the counterintuitive surface charge amplification.
The conclusions are mainly twofold:

(a) structural and charge reversal behaviour are generally well described by the various MPB
theories compared to MC simulations. For the 1:3, 3:1 systems, the theories are poor relative
to the 1:2, 2:1 cases, which is not unexpected at the higher multi-valence situations. Furthermore,
the theoretical predictions tend to be closer to the simulation data for monovalent counterions
rather than for multivalent counterions, a feature that is well known in the double layer literature.

(b) The new formulation of $F$ allows for different choices to be made for $F$. One of the new $F$'s analyzed
gives reasonable results that in some situations are at a par with MPB5. However, there are
issues, the main one being the appearance of an nonphysical shoulder in the singlet distribution
functions at lower concentrations, similar to that seen in MPB5.

	The shoulders are artifacts of the theory, which come from discontinuities in the slope
of the fluctuation potential $\phi $ at $x = 3a/2$. In the analysis outlined, although
$\phi $ is continuous across the planar surface of the truncated exclusion sphere
(cf. section~\ref{Model}), the normal derivative  is not continuous. The continuity of the normal
derivative may well be achieved by an appropriate choice of $B$, $B^*$, $C$ in equations~\ref{eq16}, \ref{eq19}.
It is to be seen if such choices in the present approach will give a simple means of eliminating
the shoulder in the distribution function. Further analytic work along the lines of MPB5 may be
difficult, so that eventually the best approach may well be a numerical evaluation of the
fluctuation potential.

\vspace{-2mm}
\section*{Acknowledgements}

   We would like to thank Dr. Z.-Y.~Wang of the School of Optoelectronic Information, Chongqing
University of Technology, People's Republic of China, for making available to us the numerical
data from his Monte Carlo simulations.

\vspace{-2mm}
\section*{Appendix}

   Evaluation of the integrals in equations~(\ref{eq13}), (\ref{eq14}) using the expressions~(\ref{eq11}) and (\ref{eq12}) for
the fluctuation potential $\phi _{i}$ in $a/2 < x <3a/2$
\begin{align}
&\int_{S_{\text I}}\phi \rd S  = \piup \left\{B(2x+a)\re^{-y}+\frac{aB^{*}}{\kappa x}\left[\re^{-\kappa s}-\re^{-\kappa (2x+a)}\right]\right\}, \\
&\int_{S_{\text W}}\frac{\partial \phi}{\partial n}\rd S  = \piup C\left[\frac{2x-3a}{a}+\frac{2f}{s}\left(s-x-\frac{a}{2}\right)\right], \\
&\int_{S_{\text W}}\left[\frac{1}{r}\frac{\partial \phi}{\partial n}-\phi \frac{\partial}{\partial n}\left(\frac{1}{r}\right)\right]\rd S  = \frac{\piup Cf}{xas}\left(a^{2}-2x^{2}+ax+as\right), \\
&\int_{S_{\text I}}\frac{\partial \phi}{\partial n}\rd S  = \piup \left\{-\frac{B}{a}(1+y)(2x+a)\re^{-y}+\left(\frac{B^{*}}{\kappa x}\right)\left[(1+y)\re^{-\kappa (2x+a)}-\left(\frac{1}{a}\right)(a+ys-\kappa sx)\re^{-\kappa s}\right]\right\}
\end{align}
and in $x > 3a/2$
\begin{align}
&\int_{S}\frac{\partial \phi}{\partial n}\rd S  = 2\piup \left[-2B(1+y)\re^{-y}+\left(\frac{B^{*}\re^{-2\kappa x}}{\kappa x}\right)(y\cosh y-\sinh y)\right], \\
&\int_{S}\phi \rd S  = 2\piup a\left[2B\re^{-y}+\left(\frac{B^{*}\re^{-2\kappa x}}{\kappa x}\right)\sinh y\right].
\end{align}

   The mean electrostatic potential integrals are
\begin{align}
&\int_{V}\nabla ^{2}\psi \rd V  = 2\piup a\Bigg[\psi (x+a)+\psi (x-a)-\frac{1}{a}\int_{x-a}^{x+a}\psi (X)\rd X\Bigg], 
\end{align}
\begin{align}
&\int_{V}\frac{\nabla ^{2}\psi}{r}\rd V  = 2\piup \left[\psi (x+a)+\psi (x-a)-2\psi (x)\right],
\end{align}
which for a truncated sphere are to be evaluated in conjunction with the linear result in $0 < x <a/2$,
\begin{equation}
\psi (x)=\psi (0)+x\left[\frac{\rd\psi}{\rd x}\right]_{x=0}.
\end{equation}

\newpage
\ukrainianpart
\title{Зміна знаку і збільшення поверхневої густини заряду у валентно-асиметричній обмеженій примітивній моделі планарних 
електричних подвійних шарів в рамках модифікованої теорії Пуасона-Больцмана}

\author{Л.Б. Буян\refaddr{label1}, К.У. Аутвейт\refaddr{label2}}

\addresses{
\addr{label1} Лабораторія теоретичної фізики, відділ фізики, А/с 70377, Університет Пуерто-Ріко, Сан Хуан, Пуерто-Ріко
\addr{label2} Відділ прикладної математики, університет Шеффілда, Шеффілд S3 7RH, Велика Британія 
}

\makeukrtitle

\begin{abstract}

В роботі переглядається і представляється у новому, більш широкому світлі модифікована теорія Пуасона-Больцмана (МТПБ) 
обмеженої примітивної моделі подвійного шару. На основі попередніх формулювань МТПБ4 і МТПБ5 виводяться відповідні рівняння, 
які дозволяють краще зрозуміти зв'язок між цими формулюваннями. МТПБ4, МТПБ5 і нове формулювання теорії застосовуються для 
аналізу структури і явища зміни знаку заряду в електролітах з асиметричною валентністю 2:1/1:2. Вивчається також явище 
індукованого поляризацією збільшення поверхневої густини заряду у системах 3:1/1:3. Проводиться порівняння з даними
моделювання методом Монте-Карло. Показано, що наведені теорії відтворюють дані моделювання з різним ступенем точності 
від якісного до майже кількісного. Результати нового варіанту теорії в багатьох випадках практично не відрізняються
від результатів МТПБ5. Однак при певних умовах, зокрема при низьких концентраціях електролітів, спостерігаються теоретичні 
артефакти у формі нефізичних ``плечей'' в унарних функціях розподілу іонів.  

\keywords електричний подвійний шар, відображення, обмежена примітивна модель, модифікована теорія Пуасона-Больцмана, 
моделювання Монте-Карло

\end{abstract}


\begin{thebibliography}{99}

\bibitem{belloni} Belloni L., J. Phys.: Condens. Matter, 2000, {\bf 12}, R549, \doi{10.1088/0953-8984/12/46/201}.
\bibitem{levin} Levin Y., Rep. Prog. Phys., 2002, {\bf 65}, 1577, \doi{10.1088/0034-4885/65/11/201}.
\bibitem{grosberg} Grosberg A.Yu., Nguyen T.T., Shklovskii B.I., Rev. Mod. Phys., 2002, {\bf 74}, 329, \doi{10.1103/RevModPhys.74.329}.
\bibitem{chertsvy} Cherstvy A.G., Phys. Chem. Chem. Phys., 2011, {\bf 13}, 9942, \doi{10.1039/c0cp02796k}.
\bibitem{bazant} Bazant M.Z., Kilic M.S., Storey B.D., Ajdari A., Adv. Colloid Interface Sci.,
2009, {\bf 152}, 48,\\ \doi{10.1016/j.cis.2009.10.001}.
\bibitem{pilon} Pilon L., Wang H., d'Entremont A., J. Electrochem. Soc., 2015, {\bf 162}, A5158, \doi{10.1149/2.0211505jes}.
\bibitem{fedorov} Fedorov M.V., Kornyshev A.A., Chem. Rev., 2014, {\bf 114}, 2978, \doi{10.1021/cr400374x}.
\bibitem{henderson} Henderson D., Boda D., Phys. Chem. Chem. Phys., 2009, {\bf 11}, 3822, \doi{10.1039/b815946g}.
\bibitem{sotres} Sotres J., Bar\'{o} A.M., Biophys. J., 2010, {\bf 98}, 1995, \doi{10.1016/j.bpj.2009.12.4330}.
\bibitem{maekawa} Maekawa Y., Shibuta Y., Sakata T., Chem. Phys. Lett., 2015, {\bf 619}, 152, \doi{10.1016/j.cplett.2014.11.068}.
\bibitem{honary1} Honary S., Zahir F., Trop. J. Pharm. Res., 2013, {\bf 12}, 255, \doi{10.4314/tjpr.v12i2.19}.
\bibitem{honary2} Honary S., Zahir F., Trop. J. Pharm. Res., 2013, {\bf 12}, 265, \doi{10.4314/tjpr.v12i2.20}.
\bibitem{galinski} Gali\'{n}ski M., Lewandowski A., St\k{e}pniak I., Electrochim. Acta, 2006, {\bf 51}, 5567,\\ \doi{10.1016/j.electacta.2006.03.016}.
\bibitem{gouy1} Gouy G., J. Phys. Theor. Appl., 1910, {\bf 9}, 457, \doi{10.1051/jphystap:019100090045700}.
\bibitem{chapman} Chapman D.L., Philos. Mag., 1913, {\bf 25}, 475, \doi{10.1080/14786440408634187}.
\bibitem{stern} Stern O., Z. Elektrochem., 1924, {\bf 30}, 508.
\bibitem{torrie1} Torrie G.M., Valleau J.P., J. Chem. Phys., 1980, {\bf 73}, 5807, \doi{10.1063/1.440065}.
\bibitem{torrie2} Torrie G.M., Valleau J.P., J. Phys. Chem., 1982, {\bf 86}, 3251, \doi{10.1021/j100213a035}.
\bibitem{snook} Snook I., van Megen W., J. Chem. Phys., 1981, {\bf 75}, 4104, \doi{10.1063/1.442571}.
\bibitem{carnie} Carnie S.L., Torrie G.M., In: Advances in Chemical Physics, Vol. 56,  Prigogine I., Rice S.A. (Eds.), John Wiley~\& Sons, Inc., Hoboken,  1984, \doi{10.1002/9780470142806.ch2}.
\bibitem{greberg} Greberg H., Kjellander R., J. Chem. Phys., 1998, {\bf 108}, 2940, \doi{10.1063/1.475681}.
\bibitem{quesada-perez1} Quesada-P\'erez M., Gonz\'alez-Tovar E., Mart\'in-Molina A., Lozada-Cassou M.,
Hidalgo-\'Alvarez R.,\\ ChemPhysChem, 2003, {\bf 4}, 234, \doi{10.1002/cphc.200390040}.
\bibitem{lyklema} Lyklema J., Colloids Surf. A, 2006, {\bf 291}, 3, \doi{10.1016/j.colsurfa.2006.06.043}.
\bibitem{lozada-cassou} Lozada-Cassou M., Gonz\'alez-Tovar E., J. Colloid Interface Sci., 2001, {\bf 239}, 285, \doi{10.1006/jcis.2001.7680}.
\bibitem{martin-molina1} Mart\'in-Molina A., Quesada-P\'erez M., Galisteo-Gonz\'alez F., Hidalgo-\'Alvarez R.,
J. Chem. Phys., 2003, {\bf 118}, 4183, \doi{10.1063/1.1540631}.
\bibitem{semenov} Semenov I., Raafatnia S., Sega M., Lobaskin V., Holm C., Kremer F.,
Phys. Rev. E, 2013, {\bf 87}, 022302, \doi{10.1103/PhysRevE.87.022302}.
\bibitem{martin-molina2} Mart\'in-Molina A., Maroto-Centeno J.A., Hidalgo-\'Alvarez R., Quesada-P\'erez M.,
Colloids Surf. A, 2008, {\bf 319}, 103, \doi{10.1016/j.colsurfa.2007.09.041}.
\bibitem{barrios} Barrios-Contreras E.A., Gonz\'{a}lez-Tovar E., Guerrero-Garc\'{\i}a G.I.,
Mol. Phys., 2015, {\bf 113}, 1190,\\ \doi{10.1080/00268976.2015.1018853}.
\bibitem{gonzalez-tovar} Gonz\'alez-Tovar E., Bhuiyan L.B., Outhwaite C.W., Lozada-Cassou M., 
J. Mol. Liq., 2017, {\bf 228}, 160, \doi{10.1016/j.molliq.2016.10.025}. 
\bibitem{jimenez-angeles} Jim\'enez-\'Angeles F., Lozada-Cassou M., J. Phys. Chem., 2004, {\bf 108}, 7286, \doi{10.1021/jp036464b}.
\bibitem{guerrero-garcia1} Guerrero-Garc\'ia G.I., Gonz\'{a}lez-Tovar E., Ch\'{a}vez-P\'{a}ez M., Lozada-Cassou M., J. Chem. Phys., 2010, {\bf 132}, 054903, \doi{10.1063/1.3294555}.
\bibitem{quesada-perez2} Quesada-P\'{e}rez M., Mart\'{\i}n-Molina A., Hidalgo-\'{A}lvarez R.,
    Langmuir, 2005, {\bf 21}, 9231, \doi{10.1021/la0505925}.
\bibitem{patra} Patra C.N., J. Chem. Phys., 2014, {\bf 141}, 184702, \doi{10.1063/1.4901217}.
\bibitem{wang1} Wang Z.-Y., Ma Y.-Q., J. Chem. Phys., 2010, {\bf 133}, 064704, \doi{10.1063/1.3469795}.
\bibitem{guerrero-garcia2} Guerrero Garcia G.I., de la Cruz M.O., J. Phys. Chem. B, 2014, {\bf 118}, 8854, \doi{/10.1021/jp5045173}.
\bibitem{wang2} Wang Z.-Y., Ma Y.-Q., Phys. Rev. E, 2012, {\bf 85}, 062501, \doi{10.1103/PhysRevE.85.062501}.
\bibitem{wang3} Wang Z.-Y.,  J. Stat. Mech.: Theory Exp., 2016, {\bf 2016}, 043205, \doi{10.1088/1742-5468/2016/04/043205}.
\bibitem{outh1} Outhwaite C.W., Bhuiyan L.B., Levine S., J. Chem. Soc., Faraday Trans. 2,
1980, {\bf 76}, 1388,\\ \doi{10.1039/F29807601388}.
\bibitem{outh2} Outhwaite C.W., Bhuiyan L.B.,  J. Chem. Soc., Faraday Trans. 2,
1982, {\bf 78}, 775, \doi{10.1039/F29827800775}.
\bibitem{outh3} Outhwaite C.W., Bhuiyan L.B., J. Chem. Soc., Faraday Trans. 2,
1983, {\bf 79}, 707, \doi{10.1039/F29837900707}.
\bibitem{kirkwood} Kirkwood J.G., J. Chem. Phys., 1934,
{\bf 2}, 767, \doi{10.1063/1.1749393}.
\bibitem{loeb} Loeb A.L., J. Colloid Sci., 1951, {\bf 6}, 75, \doi{10.1016/0095-8522(51)90027-X}.
\bibitem{outh5} Outhwaite C.W., Bhuiyan L.B., Mol. Phys., 2014, {\bf 112}, 2963, \doi{10.1080/00268976.2014.922706}.
\bibitem{outh6} Outhwaite C.W., Bhuiyan L.B., Levine S., Chem. Phys. Lett., 1981, {\bf 78}, 413,\\ \doi{10.1016/0009-2614(81)85226-8}.
\bibitem{bhuiyan1} Bhuiyan L.B., Outhwaite C.W., Levine S., Mol. Phys., 1981, {\bf 42}, 1271, \doi{10.1080/00268978100100961}.
\bibitem{levine1} Levine S., Outhwaite C.W., J. Chem. Soc., Faraday Trans. 2, 1978, {\bf 74}, 1670, \doi{10.1039/f29787401670}, [J.~Chem. Soc., Faraday Trans. 2, 1980, {\bf 76}, 221 (Corrigendum), \doi{10.1039/f29807600221}].
\bibitem{levine2} Levine S., Outhwaite C.W., Bhuiyan L.B., J. Electroanal. Chem., 1981, {\bf 123}, 105,\\ \doi{10.1016/s0022-0728(81)80046-0}.
\bibitem{outh7} Outhwaite C.W., Lamperski S., Condens. Matter Phys., 2001, {\bf 4}, 739, \doi{10.5488/CMP.4.4.739}.
\bibitem{fischer} Fischer J., Mol. Phys., 1977, {\bf 33}, 75, \doi{10.1080/00268977700103061}.
\bibitem{metropolis} Metropolis N., Ulam S., J. Am. Stat. Assoc., 1949, {\bf 44}, 335, \doi{10.1080/01621459.1949.10483310}.
\bibitem{allen} Allen M.P., Tildsley D., Computer Simulation of Liquids, Oxford University Press, New York, 1989.
\bibitem{bhuiyan2} Bhuiyan L.B., Outhwaite C.W., Henderson D., Alawneh M., Mol. Phys., 2007, {\bf 105}, 1395,\\ \doi{10.1080/00268970701355795}.
\bibitem{bhuiyan3} Bhuiyan L.B., Outhwaite C.W., Henderson D., J. Chem. Eng. Data, 2011, {\bf 56}, 4556, \doi{10.1021/je2005193}.
\bibitem{boda} Boda D., Chan K.-Y., Henderson D., J. Chem. Phys., 1998, {\bf 109}, 7362, \doi{10.1063/1.477342}.
\bibitem{torrie3} Torrie G.M., Valleau J.P., Patey G.N., J. Chem. Phys., 1982, {\bf 76}, 4615, \doi{10.1063/1.443541}.
\bibitem{bellman} Bellman R., Kalaba R., Quasilinearization and Nonlinear Boundary Value Problems,
Elsevier, New York, 1965.
\bibitem{outh8a} Outhwaite C.W., Chem. Phys. Lett., 1970, {\bf 7}, 636, \doi{10.1016/0009-2614(70)87027-0}.
\bibitem{ulander} Ulander J., Kjellander R., J. Chem. Phys., 2001, {\bf 114}, 4893, \doi{10.1063/1.1350449}.
\bibitem{bhuiyan4} Bhuiyan L.B., Outhwaite C.W., Phys. Chem. Chem. Phys., 2004, {\bf 6}, 3467, \doi{10.1039/b316098j}.
\bibitem{henderson2} Henderson D., Blum L., J. Chem. Phys., 1978, {\bf 69},
5441, \doi{10.1063/1.436535}.
\bibitem{henderson3} Henderson D., Blum L., Lebowitz J.L., J. Electroanal. Chem., 1979, {\bf 102},
315,\\ \doi{10.1016/S0022-0728(79)80459-3}.
\bibitem{martin} Martin Ph.A.,  Rev. Mod. Phys., 1988, {\bf 60}, 1075, \doi{10.1103/RevModPhys.60.1075}.
\bibitem{outh8} Outhwaite C.W., Bhuiyan L.B., J. Chem. Phys., 1986, {\bf 85},
4206, \doi{10.1063/1.451812}.
\bibitem{bratko} Bratko D., Bhuiyan L.B., Outhwaite C.W., J. Phys. Chem., 1986, {\bf 90},
6248, \doi{10.1021/j100281a036}.
\bibitem{silvestre} Silvestre-Alcantara W., Bhuiyan L.B., Outhwaite C.W., Henderson D.,
Collect. Czech. Chem. Commun., 2010, {\bf 75}, 425, \doi{10.1135/cccc2009098}.
\end{thebibliography}
\end{document}